\documentclass[aps,prl,twocolumn,groupedaddress]{revtex4}
\usepackage{amsmath}
\usepackage{graphicx}
\usepackage{color}

\begin{document}

\title{Optical Absorption by Dirac Excitons\\ in Single-Layer Transition-Metal Dichalcogenides}

\author{Maxim Trushin$^1$}
\author{Mark Oliver Goerbig$^2$}
\author{Wolfgang Belzig$^1$}
\affiliation{$^1$University of Konstanz, Fachbereich Physik, M703 D-78457 Konstanz, Germany}
\affiliation{$^2$Laboratoire de Physique des Solides, Univ. Paris-Sud, Universit\'e Paris-Saclay, CNRS UMR 8502, F-91405 Orsay, France}

\date{\today}

\begin{abstract}
We develop an analytically solvable model able to qualitatively explain nonhydrogenic exciton spectra observed recently 
in two-dimensional (2d) semiconducting transition metal dichalcogenides.
Our exciton Hamiltonian explicitly includes additional angular momentum associated with the pseudospin degree of freedom
unavoidable in 2d semiconducting materials with honeycomb structure.
We claim that this is the key ingredient for understanding the nonhydrogenic exciton spectra that was missing so far.
\end{abstract}


\maketitle

{\em Introduction. ---}
Following the discovery of graphene \cite{Geim2011}, two-dimensional (2d) materials have
experienced a boom over the last decade \cite{Xu2013}.
One of their most prominent representatives are transition metal dichalcogenides (TMDs) with the stochiometric formula
MX$_2$, where M represents a transition metal, like Mo or W, and X stands for a
chalcogenide (S, Se, or Te) \cite{Mak2010}.
In contrast to the parent bulk crystals TMD monolayers are direct bandgap semiconductors \cite{Korn2011,Splendiani2010} with a bandgap
in the visible spectrum. Practical applications of 2d TMDs are already envisaged \cite{Nanoscale2015roadmap} 
with the emphasis in optoelectronics \cite{Wang2012} and photodetection \cite{Natnano2014review}, where optical absorption plays a central role.
It is therefore of utmost importance to understand the dominating optical absorption mechanism in 2d TMDs,
which has strong excitonic character 
\cite{NatMat2015xavier,PRL2014chernikov,PRL2014he,SSC2015hanbicki,SR2015zhu,NatMat2014ugeda,Nature2014ye,PRL2015wang,PRB2013mitioglu,2DM2015wang}.

\begin{figure}
\includegraphics[width=\columnwidth]{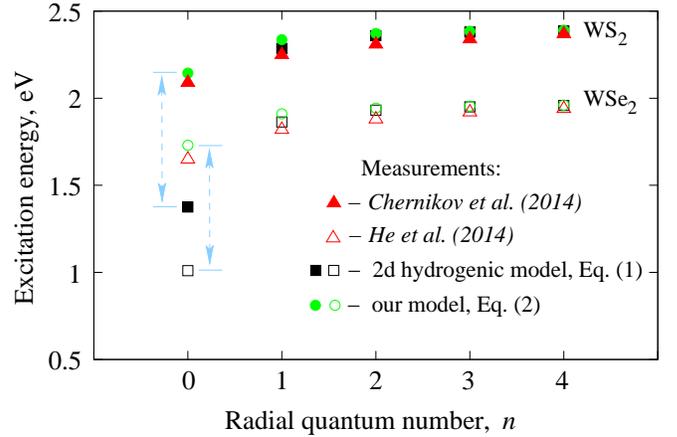}
\caption{\label{fig1} (Color online) Exciton spectrum for some 2d TMDs: theory and measurements. The environment is SiO$_2$ in both cases with the dielectric constant $\epsilon=3.9$,
i.e. the interaction constant $\alpha=e^2/(\sqrt{2} \epsilon \hbar v)$ is determined solely by the band parameter $v$.
Only s-states with $m=0$ in Eq.~(\ref{Rydberg}) or with $j=\pm 1/2$ in Eq.~(\ref{main}) are optically active.
The measurements for WS$_2$ and WSe$_2$ are taken from Refs. \cite{PRL2014chernikov} and \cite{PRL2014he} respectively.
The calculations have been performed for WS$_2$ with the bandgap $\Delta=2.4$eV, $\hbar v = 1.25 \mathrm{eV} \times 3.197$\AA~ \cite{Kim2016}
and resulting interaction parameter $\alpha=0.65$, and for WSe$_2$ we take $\Delta=1.97$eV, $\hbar v = 1.13 \mathrm{eV} \times 3.310$\AA~ \cite{Kim2016}
with resulting $\alpha=0.7$. The arrows highlight the difference in binding energies evaluated from  Eq.~(\ref{Rydberg}) and Eq.~(\ref{main}). }
\end{figure}

An exciton is a bound state of an electron and a hole which are attracted to each other by the Coulomb force \cite{Elliott1957}.
The electron-hole (e-h) pair in 2d semiconductors has usually been described as a 2d hydrogen-like system with the reduced mass $\mu^{-1}=m_e^{-1}+ m_h^{-1}$
and the excitation spectrum \cite{VaskoBook}
\begin{equation}
\label{Rydberg}
 E_{nm}=\Delta -\frac{e^4 \mu}{2\epsilon^2 \hbar^2} \frac{1}{(n+|m|+1/2)^2},
\end{equation}
where $e$ is the elementary charge, $\epsilon$ is the dielectric constant, $\hbar$ is the Planck constant, 
and $n=0,1,2...$, $m=0,\pm 1, \pm 2$ are the radial and magnetic quantum numbers, respectively.
The fundamental bandgap $\Delta$ is effectively reduced by the binding energy $E_b= 2e^4 \mu/(\epsilon \hbar)^2$.
However, the exciton spectrum in 2d TMDs \cite{PRL2014chernikov,PRL2014he,NanoLett2015hill} does not resemble the conventional Rydberg series (\ref{Rydberg}).
A few previous attempts to solve the problem involve non-Coulomb interactions \cite{PRL2016olsen,PRB2015berkelbach,PRB2015wu,PRB2014berghauser}, 
exciton p-states \cite{JoP2015stroucken}, Berry phase \cite{PRL2015zhou,PRL2015srivastava}, and multiple ab-initio and other numerical calculations 
\cite{Nature2014ye,PRL2013qiu,PRB2012ramasubramaniam,PRB2012komsa,PRB2013shi,PRB2012cheiwchanchamnangij,PRB2016echeverry,2DM2015wang}.
Despite extensive theoretical efforts, a simple analytical model that provides insight into the exciton problem is still missing.
In this paper we show that a proper model has to account for the interband coupling between electron and hole states, the strength of which has been quantified
as ``Diracness'' \cite{EPL2014goerbig},
inherited from the single-particle effective Hamiltonian for carriers in TMDs \cite{PRL2012xiao}. We find that the excitation spectrum for the experimentally relevant
regime $\Delta\gg E_b$ (shallow bound states) reads
\begin{equation}
 \label{main}
 E_{nj} = \Delta - \frac{e^4 \mu}{2\epsilon^2 \hbar^2} \frac{1}{(n+|j|+1/2)^2},
\end{equation}
where $j=m+1/2$ is the {\em total} (i.e. orbital and pseudospin \cite{PRL2012xiao}) angular momentum.
Here, the binding energy is $E_b=e^4 \mu/(2\epsilon^2 \hbar^2)$.
As compared with Eq.~(\ref{Rydberg}), this spectrum shows much better agreement
with the measurements \cite{PRL2014chernikov,PRL2014he}, Fig.~\ref{fig1}, 
and, along with the effective Hamiltonian (\ref{mainH}), represents our main finding.

{\em Effective exciton Hamiltonian. ---}
The rigorous derivation of an exciton Hamiltonian for 2d TMDs involves coupling of two massive Dirac particles \cite{PRA2013berman,JPCM2015li,PRB2012berman-a,PRB2012berman-b,PRB2008berman}
that is not analytically tractable even in the limit of zero mass \cite{PRB2010sabio,EPL2013mahmoodian,EPJB2012groenqvist}. 
We therefore derive an {\em effective} exciton Hamiltonian
that is inspired by the one-particle Hamiltonian for a given valley and spin \cite{PRL2012xiao},
\begin{equation}
\label{1p}
H_1 = \left(\begin{array}{cc} \Delta/2  & \hbar v k \mathrm{e}^{-i \theta} \\
\hbar v k \mathrm{e}^{i \theta} & -\Delta/2
\end{array}\right),
\end{equation}
where $\tan\theta=k_y/k_x$, and $v$ is the velocity parameter which can be either measured \cite{Kim2016} or calculated \cite{PRL2012xiao}.
This Hamiltonian already contains an interband coupling via off-diagonal terms, such that the electron and hole states are not independent even
without Coulomb interactions. This observation alone suggests
the possibility that atypical quantum effects can play a role in bound states.

Let us now derive an effective-mass model that takes into account the pseudospin degree of freedom, which makes the electron and hole states entangled
via off-diagonal terms in $H_1$.
The eigenvalues of (\ref{1p}) are $\pm\sqrt{(\hbar v k)^2+\Delta^2/4}$, which in parabolic approximation
suggest the same effective mass $m_{e,h}=\Delta/(2v^2)$ for electrons and holes. The excitonic reduced effective mass
should therefore be $\mu=\Delta/(4v^2)$ with the bound state spectrum given by Eq.~(\ref{Rydberg}).
A somewhat more comprehensive parametrization of the single-particle Hamiltonian \cite{EPL2014goerbig,2DM2015kormanyos}, 
$\Delta/2 \rightarrow \Delta/2 + \hbar^2k^2/2m_0$, 
just leads to the renormalization of the exciton mass $1/\mu=4v^2/\Delta + 2/m_0$.

We assume that the center of mass does not move for optically excited e-h pair \cite{Elliott1957}, and
the electron and hole momenta have the same absolute values but opposite directions.
The two-particle Hamiltonian without Coulomb interactions is therefore given by the the tensor product \cite{PRB2013rodin}
$H_2=H_1 \otimes I_2 - I_2 \otimes (T H_1 T^{-1})$ (here $I_2$ is the $2\times 2$ unit matrix, and $T H_1 T^{-1}$ is the time reversal of $H_1$), and reads
\begin{equation}
\label{2p}
H_2=\left( \begin{array}{cccc}
 0 & \hbar k v \mathrm{e}^{i \theta }  & \hbar k v \mathrm{e}^{-i \theta }  & 0 \\
\hbar k v \mathrm{e}^{-i \theta }  & \Delta & 0 & \hbar k v \mathrm{e}^{-i \theta }  \\
 \hbar k v \mathrm{e}^{i \theta }  & 0 & -\Delta & \hbar k v \mathrm{e}^{i \theta }  \\
 0 & \hbar k v \mathrm{e}^{i \theta }  & \hbar k v \mathrm{e}^{-i \theta }  & 0 \\
\end{array}
\right).
\end{equation}
The Hamiltonian $H_2$ has four eigenvalues: $E_{1,4}=\pm 2\sqrt{\hbar^2 v^2 k^2 +\Delta^2/4}$, $E_{2,3}=0$ depicted in Fig.~\ref{fig2}a
with their physical meaning explained in Fig.~\ref{fig2}b. Using the transformation $P^{-1} H_2 P$ with 
\begin{equation}\label{trafo}
P=\left( \begin{array}{cccc}
 \cos\frac{\Theta}{2} & 0 & \sin\frac{\Theta}{2} & 0 \\
 0 & \cos\frac{\Theta}{2} & 0 & \sin\frac{\Theta}{2} \\
 \mathrm{e}^{i \theta } \sin\frac{\Theta}{2} & 0 & -\mathrm{e}^{i \theta } \cos\frac{\Theta}{2} & 0 \\
 0 & \mathrm{e}^{i \theta } \sin\frac{\Theta}{2} & 0 & -\mathrm{e}^{i \theta} \cos\frac{\Theta}{2} \\
\end{array} \right)
\end{equation}
and $\tan\Theta=2\hbar v k/\Delta$, Eq.~(\ref{2p}) can be block-diagonalized into a matrix $H_2=H_2^+\oplus H_2^-$
with 
\begin{equation}
\label{H2pm}
H_2^\pm = 
\left(\begin{array}{cc}
-\frac{\Delta}{2}\pm\sqrt{\hbar^2 v^2 k^2 +\frac{\Delta^2}{4}} & \hbar v k \mathrm{e}^{i \theta}  \\
 \hbar v k \mathrm{e}^{-i \theta}  &  \frac{\Delta}{2}\pm\sqrt{\hbar^2 v^2 k^2 +\frac{\Delta^2}{4}}  \\
\end{array}
\right).
\end{equation}
$H_2^+$ and $H_2^-$ have the eigenvalues $E_{1,3}$ and $E_{2,4}$ correspondingly with only the former describing the excitons we are interested in.
The diagonal terms in (\ref{H2pm}) can be written within the effective mass approximation, but
the matrix remains in the peculiar mixed ``Dirac-Schr\"odinger'' form: The off-diagonal ``Dirac'' terms $ \hbar v k \mathrm{e}^{\pm i \theta}$
couple the ``Schr\"odinger'' states. Our goal is to write an effective-mass Hamiltonian which mimics this feature,
but remains tractable at the analytical level.
The minimal Hamiltonian which fulfills these criteria reads
\begin{equation}
\label{eff}
H_2^\mathrm{eff}= \left(\begin{array}{cc}  \frac{\hbar^2 k^2}{2 \mu} & \hbar k \sqrt{\frac{\Delta}{2\mu}} \mathrm{e}^{-i \theta} \\
\hbar k \sqrt{\frac{\Delta}{2\mu}} \mathrm{e}^{i \theta} & \Delta
\end{array}\right).
\end{equation}
There is only a single parabolic branch $E_k=\Delta + \frac{\hbar^2 k^2 }{2 \mu}$ in the spectra of $H_2^\mathrm{eff}$, see Fig.~\ref{fig2}c.
The other branch $E_0=0$ is dispersionless, as it should follow from the more rigorous model Hamiltonian $H_2$. 
The remaining task is to switch on the Coulomb interaction $V(r)=-e^2/\epsilon r$ and change the momenta to the corresponding operators.
Using notations of Eq.~(\ref{1p}) we write the resulting Hamiltonian as
\begin{equation}
\label{mainH}
\hat H= \left(\begin{array}{cc}  \frac{2\hbar^2 v^2 }{\Delta}(\hat k_x^2 + \hat k_y^2) + V(r) &  \sqrt{2}\hbar v (\hat k_x - i \hat k_y) \\
 \sqrt{2}\hbar v (\hat k_x + i \hat k_y)  & \Delta+V(r)
\end{array}\right),
\end{equation}
which contains no pseudo-differential operators, such as $\sqrt{\Delta^2/4 + \hbar^2 v^2 \hat  k^2}$ that we
would have to deal with starting directly from Eq.~(\ref{H2pm}). 
Before solving the Hamiltonian in a strictly quantum-mechanical manner, we should keep in mind the following approximations
involved. Our model (\ref{mainH}) has not been obtained directly from the original $4\times 4$ model (\ref{2p}) of coupled 2d Dirac fermions --- indeed, the decoupling
transformation ({\ref{trafo}) depends on the lattice momentum $k$ and therefore does not commute with the potential $V(r)$. Strictly speaking, the transformation 
would generate corrective terms that are neglected in the present treatment. Furthermore, as a consequence of the underlying relativistic structure of Dirac fermions, 
the relative momentum $k$ is not decoupled from the center-of-mass momentum of the exciton, and our treatment is thus valid only in the exciton rest frame. However, 
(relativistic) corrective terms are expected to be small in experimentally relevant situations due to the rather large gap $\Delta$ in 2d TMDs. The major merit
of our model (\ref{mainH}) is to reproduce the relevant excitonic bands while 
retaining the off-diagonal terms whose manifestation we are investigating here.

\begin{figure}
\includegraphics[width=\columnwidth]{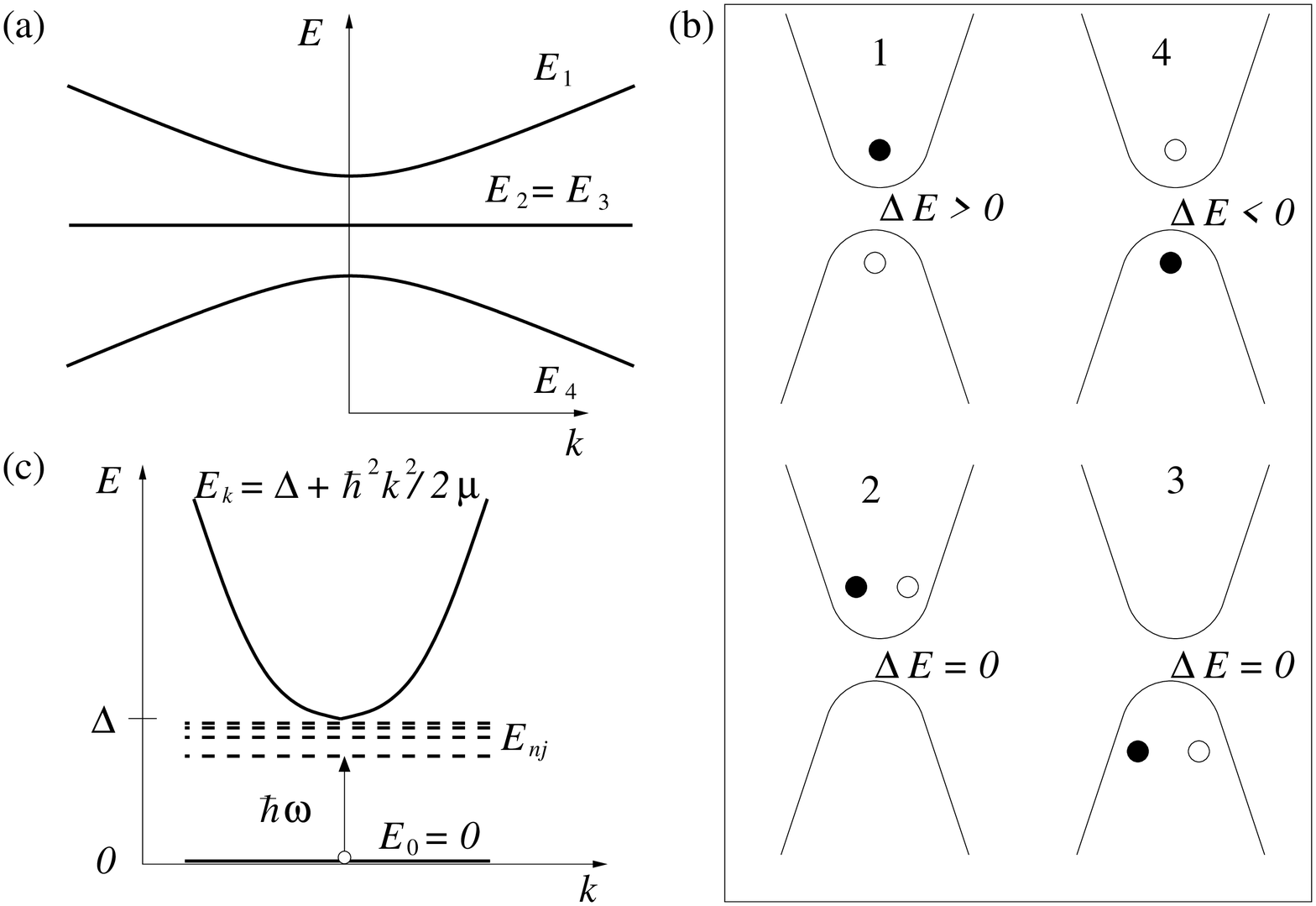}
\caption{\label{fig2} (a) Four branches in the spectrum of 
non-interacting two-particle Hamiltonian (\ref{2p}).
(b) Possible configurations of two particles corresponding to these four branches. 
The first configuration with the positive energy is an exciton. The fourth configuration corresponds
to the opposite situation where the electron remains in the valence band, and the hole is in the conduction band.
There are two possible configurations with zero energy, where both particles are in the same band.
(c) The two branches of the effective exciton Hamiltonian (\ref{eff}) shown by solid curves and
the bound states (\ref{main}) depicted by the dashed lines. One of possible optical transitions is shown
for a given radiation frequency $\omega$. Note that only the states with $j=\pm 1/2$ in Eq.~(\ref{main}) are optically active in one-photon processes.}
\end{figure}

{\em Excitonic spectrum. ---}
In order to solve the spectral problem $\hat H \Psi_{nm} = E_{nm} \Psi_{nm}$ we employ polar coordinates $\{ \varphi, r\}$ and 
define the following dimensionless quantities
$$
\varepsilon = \frac{E}{\Delta}, \quad \lambda=\sqrt{1-\varepsilon}, \quad  \rho= \lambda\frac{\sqrt{2}\Delta}{\hbar v}  r, \quad 
\alpha=\frac{e^2}{\sqrt{2} \epsilon \hbar v}.
$$
Note that $\varepsilon<1$ because  we are interested in the bound states with the energies below $\Delta$.
We look for the solution in the form
\begin{equation}
\label{psinm}
\Psi_{nm} = \rho^{|m|}\left(\begin{array}{c}  F(\rho) \mathrm{e}^{i m \varphi} \\ i  G(\rho) \mathrm{e}^{i (m+1) \varphi}
\end{array}\right),
\end{equation}
and the equations for the radial parts read
\begin{eqnarray}
\nonumber && 
\frac{\partial^2 F }{\partial \rho^2}  + \frac{2|m|+1}{\rho}\frac{\partial F}{\partial \rho}
-\frac{1}{2\lambda}\left(\frac{\partial G}{\partial \rho} + \frac{|m|+m+1}{\rho} G \right) \\
\label{eq1}
&& + \left( \frac{\alpha}{2\lambda \rho} + \frac{1}{4\lambda^2} - \frac{1}{4} \right) F =0,\\
&& \frac{|m|-m}{\rho} F + \frac{\partial F}{\partial \rho} + \frac{2\alpha-\lambda \rho}{2\rho}G  =0.
\label{eq2}
 \end{eqnarray}
The equations can be now decoupled easily. From Eq.~(\ref{eq2}) we obtain
\begin{equation}
G=-\frac{2 \rho}{2 \alpha  -\lambda \rho }\left( \frac{\partial F}{\partial \rho} +\frac{|m|-m}{\rho} F\right)
\end{equation}
and the equation for $F$ reads
\begin{eqnarray}
\nonumber && \frac{\partial^2 F }{\partial \rho^2}  + \left( \frac{2|m|+1}{\rho} + f(\rho) \right) \frac{\partial F}{\partial \rho} \\
&& + \left( \frac{\alpha}{2\lambda \rho} - \frac{1}{4} + f(\rho) \frac{|m|-m}{\rho}\right) F =0,
\label{final1}
\end{eqnarray}
where 
\begin{equation}
 \label{ff}
 f(\rho)= \frac{2\alpha}{(2\alpha - \lambda \rho)[2\alpha \lambda + (1-\lambda^2)\rho]}.
\end{equation}

In the formal limit $\alpha \to \infty$ we have $G = 0$, and the $f(\rho)$-dependent terms in (\ref{final1}) are neglected.
Using the asymptotic behavior at $\rho\to \infty$ and making the substitution  $\tilde{F}= F \mathrm{e}^{-\rho/2}$
we arrive at the confluent hypergeometric equation
\begin{equation}
\label{strong}
\rho \frac{\partial^2 \tilde{F} }{\partial \rho^2} + (2|m|+1 - \rho ) \frac{\partial \tilde{F}}{\partial \rho}
+ \left( \frac{\alpha}{2\lambda} - |m| - \frac{1}{2} \right)\tilde{F} =0.
\end{equation}
The wave function must vanish at $\rho \to \infty$, thus,   $\frac{\alpha}{2\lambda} -|m| - \frac{1}{2}=n$
must be a positive integer or zero and the spectrum is given by (\ref{Rydberg}).
This result is expected  at $\alpha\to \infty$ because the diagonal (i.e. ``Schr\"odinger'') part in the Hamiltonian (\ref{mainH}) dominates in this limit.
This regime is however unphysical since the bound states cannot lie deeper than the band gap size, so that
the conventional series (\ref{Rydberg}) cannot be realized in 2d TMDs.

The opposite regime of small $\alpha$ makes the discrete levels $E_mn$ be closer to the bottom of the continuous spectral region
that results in smaller $\lambda$. This is the shallow bound states approximation relevant for excitons
because the binding energy is much smaller than the bandgap even for 2d TMDs.
Note, that $\lambda$ is of the same order as $\alpha$, thus, $\rho^2 \ll \rho$ and
$f(\rho)\approx 1/\rho$.
The asymptotic behavior of $F$ at $\rho\to\infty$ is $F=\mathrm{e}^{-\rho/2}$ and $F=\rho^\gamma$  at $\rho \to 0$, where 
$\gamma=\left|m+1/2\right| - (|m|+1/2)$.
Hence, we make the substitution $F= \tilde{F} \rho^\gamma \mathrm{e}^{-\rho/2}$, and
the resulting equation for $\tilde{F}$ reads
\begin{equation}
\label{weak}
\rho \frac{\partial^2 \tilde{F} }{\partial \rho^2} + (1+ |2m+1| - \rho ) \frac{\partial \tilde{F}}{\partial \rho}
+ \left( \frac{\alpha}{2\lambda} -\frac{1}{2} -  \frac{|2m+1|}{2} \right)\tilde{F} =0.
\end{equation}
We have arrived at the confluent hypergeometric equation again but with parameters different from Eq.~(\ref{strong}).
Indeed, the radial quantum number is now defined as
$\frac{\alpha}{2\lambda} -\frac{1}{2} -  \frac{|2m+1|}{2} =n$ with $n=0,1,2,..,$
and the energy spectrum is given by Eq.~(\ref{main}).
The corresponding eigenstates are given by the spinor (\ref{psinm}), where
$G(\rho)$ is given by a linear combination of confluent hypergeometric functions with different $n$'s,
cf. \cite{PRB2007novikov,PRB2008pereira}.

{\em Excitonic optical absorption. ---}
The model allows for analysis of the optical selection rules relevant for the measurements 
\cite{PRL2014chernikov,PRL2014he,SSC2015hanbicki,SR2015zhu,NatMat2014ugeda,Nature2014ye,PRL2015wang}.
As it follows from Fig.~(\ref{fig2})c, the optical transitions occur between the ``vacuum'' state $E_0$ and the discrete levels $E_{nj}$.
Even if the states of the continuous spectrum are influenced by the Coulomb interaction,  we assume that 
the most important symmetry features are already encoded in the unperturbed eigenstates of $\hat H$ at $V(r)=0$ which read
\begin{equation}
\label{psi0}
 \Psi^0_{km} = C\left(\begin{array}{c}   \cos\xi \, J_m(kr) \mathrm{e}^{i m \varphi} \\ -i \sin\xi \,  J_{m+1}(kr) \mathrm{e}^{i (m+1) \varphi} 
\end{array}\right),
\end{equation}
for the dispersionless branch $E_0=0$, and 
\begin{equation}
 \label{psik}
 \Psi_{km} = C\left(\begin{array}{c}   \sin\xi \, J_m(kr) \mathrm{e}^{i m \varphi} \\ i \cos\xi \,  J_{m+1}(kr) \mathrm{e}^{i (m+1) \varphi}
\end{array}\right),
\end{equation}
for the parabolic branch $E_k=\Delta + \hbar^2 k^2/(2 \mu)$.
Here, $J_m(kr)$ is the Bessel function, $\tan \xi = \sqrt{2} \hbar v k / \Delta$, and $C$ is the normalization constant.
The light-quasiparticle interaction Hamiltonian is derived from $\hat H$ substituting the quasiparticle momentum by
the vector potential describing the electromagnetic field \cite{Trushin2012}.
To simplify our analysis of the optical transitions we enforce momentum conservation, i.e. $k\to 0$ in Eqs.~(\ref{psi0},\ref{psik}).
In particular, $\Psi^0_{km} \propto \left(\delta_{m,0}, 0\right)^T$ in this limit.
The linear-response light-quasiparticle interaction Hamiltonian then reads
\begin{equation}
 \label{Hint}
 \hat H_i = \frac{ev \cal{E}}{\sqrt{2}\omega} 
 \left(\begin{array}{cc} 0 & \mathrm{e}^{-i\phi_{\cal{E}}} \\
\mathrm{e}^{i\phi_{\cal{E}}}  & 0
\end{array}\right),
\end{equation}
where $\cal{E}$ is the electromagnetic wave amplitude, $\omega$ is its frequency, and $\tan\phi_{\cal{E}} = {\cal{E}}_{y}/{\cal{E}}_{x}$
is the polarization angle.
The transition rate from the ``vacuum'' state $\Psi^0_{m'k}$ to the state $\Psi_{nm}$ can be evaluated by means
of Fermi's golden rule with the transition matrix elements $\langle \Psi_{nm} |\hat H_i | \Psi^0_{km'} \rangle$.
At $\alpha \to \infty$ the lower part of the spinor $\Psi_{nm}$ vanishes, hence, the transition rate is zero in this limit.
In the case of shallow levels the optical transitions are allowed from the ``vacuum'' with $m'=0$  to the bound state with $m=-1$ and arbitrary $n$.
Thus, the optically active series are given by (\ref{main}) with $j=-1/2$.
The opposite corner of the Brillouin zone gives the same excitation spectrum with $j=1/2$.
This constitutes the one-photon optical selection rule in the excitonic absorption.

{\em Discussion and conclusion. ---}
The spectrum (\ref{main}) is not symmetric with respect to $m$, i.e. $E_m\neq E_{-m}$ because the initial one-particle Hamiltonian (\ref{1p})
is not time-reversal invariant. To restore the time-reversal invariance we have to consider both non-equivalent corners of 
the full Brillouin zone. The
spectrum (\ref{main}) is however symmetric with respect to $j$, i.e. $E_j= E_{-j}$.
In particular, Eq.~(\ref{main}) for s-states with $j=\pm 1/2$ reproduces the standard {\em three-dimensional} hydrogen-like spectrum 
for an e-h pair with reduced effective mass $\mu$. This is the pseudospin that
removes $-1/2$ in the standard 2d hydrogen-like spectrum (\ref{Rydberg}) and represents the main feature of our model.
Despite its formal simplicity it has important consequences for the exciton binding energy and level spacing.
Indeed, the standard model (\ref{Rydberg}) overestimates the binding energy by the factor of $4$ as well as 
the level spacing between the lowest and the first excited bound states, which is $8E_b/9$ in the hydrogenic model (\ref{Rydberg}),
but only $3E_b/4$ within our model (\ref{main}).  The reduced level spacing has been experimentally observed in WS$_2$ and WSe$_2$, see e.g. Fig.4 in Ref.~\cite{PRL2014he}.
The exciton spectrum measured in MoS$_2$ allows only for an ambiguous interpretation \cite{NanoLett2015hill} due to weak spin-orbit splitting
between the A and B exciton series but it also demonstrates reduced level spacing.  
The measured excitation energy of the 2s state in the B-series ($2.24$eV \cite{NanoLett2015hill}) is overestimated by the tight-binding  ($2.27$eV \cite{PRB2015wu}),
and first-principles ($2.32$ eV\cite{PRL2013qiu}) calculations, as well as by our Eq.~(\ref{main}) resulting in $2.30$eV
at $\hbar v = 1.01 \mathrm{eV} \times 3.193$\AA~ \cite{Kim2016} and $\Delta=2.4$eV.
While Refs.\cite{PRL2013qiu,PRB2015wu} include pseudospin along with many other effects 
our model emphasizes its importance explicitly.

We do not take into account non-Coulomb interactions due to the non-local screening in thin semiconductor films \cite{Keldysh,PRL2014chernikov,PRL2016olsen},
as they are considered less important than pseudospin within our model. The non-local screening makes the dielectric constant
$\epsilon$ dependent on the exciton radius which increases with $n$ \cite{PRL2014chernikov,PRL2016olsen}, whereas
pseudospin modifies the very backbone of the exciton model --- the fundamental $1/(n+1/2)^2$ spectral series.
Indeed, our model combines the fundamental features of the standard hydrogenic model resulting in the Rydberg series (\ref{Rydberg})
and an exotic spectrum obtained for purely ``Dirac'' excitons neglecting the ``Schr\"odinger'' part  \cite{PRL2007shytov,PRB2007novikov,PRB2013rodin,JoP2015stroucken}
\begin{equation}
 \label{Dirac}
 E_{nj}= \Delta \frac{n+\sqrt{j^2 - \tilde{\alpha}^2}}{\sqrt{\tilde{\alpha}^2 + (n+ \sqrt{j^2 - \tilde{\alpha}^2})^2}},
\end{equation}
with the renormalized interaction constant $\tilde{\alpha}=e^2/(2\epsilon\hbar v)$.
Similar to our model, Eq.~(\ref{Dirac}) involves the total angular momentum $j$, but it does not lead to Eq.~(\ref{main}) even at small $\tilde{\alpha}$.
Eq.~(\ref{Dirac}) suggests the collapse of s-states at $\tilde{\alpha}>1/2$  and,  in order to fit the measurements \cite{PRL2014chernikov,PRL2014he},
p-states are employed \cite{JoP2015stroucken}. In contrast, the s-excitons never collapse in our model, in accordance with the experimental claims
\cite{PRL2014chernikov,PRL2014he,SSC2015hanbicki,SR2015zhu,NatMat2014ugeda,Nature2014ye,PRL2015wang,2DM2015wang}.
Thus, the excitons in 2dTMDs are neither Schr\"odinger nor Dirac quasiparticles but retain the properties of both, as reflected in our effective Hamiltonian (\ref{mainH}).

To conclude, we propose an analytical model for optical absorption by massive 2d Dirac excitons which explains
the origin of the peculiar excitonic spectrum in 2dTMDs. The key feature is the combination of ``Schr\"odinger'' and ``Dirac'' terms
in the effective Hamiltonian (\ref{mainH}) which follows from a more rigorous two-particle model (\ref{2p}). The model can be further employed to describe other
experiments that involve excitons in 2d TMDs, e.g.  valley-resolved  pump-probe spectroscopy in MoS$_2$ \cite{PRB2015dalconte}.

We acknowledge financial support from the Center for Applied Photonics (CAP) and thank Alexey Chernikov for discussions.

\bibliography{excitons-theory.bib,excitons-experiment.bib}

\end{document}